\documentstyle[aps,prl]{revtex}

\draft

\begin{document}


\title{Testing the correlation of ultra-high energy cosmic rays
with high redshift sources}

\author{G\"unter Sigl$^{a}$\footnote{E-mail address: {\tt
sigl@iap.fr}}, Diego F. Torres$^{b}$\footnote{E-mail address: {\tt
dtorres@venus.fisica.unlp.edu.ar}}, Luis A. Anchordoqui$^c$, and
Gustavo E. Romero$^b$
\\
{\small $^a$ Institut d'Astrophysique de Paris, UPR 341 CNRS,
98bis Boulevard Arago, 75014 Paris, France}\\ {\small
$^b$Instituto Argentino de Radioastronom\'{\i}a, C.C.5, 1894 Villa
Elisa, Buenos Aires, Argentina} \\ {\small $^c$Department of
Physics, Northeastern University, Boston, Massachusetts 02115,
USA} }

\maketitle

\begin{abstract}
We study the correlation between compact radio quasars
or 3EG gamma-ray blazars and the arrival directions of cosmic rays above
$10^{19}\,$eV using an updated list of air shower detections.
Our Monte Carlo simulations reveal no significant correlations above
random and some previous positive results appear to be an effect
of the small sample size. Consequently, unless somehow severely
deflected, there is no evidence for ultra-high energy cosmic ray
primaries being new particles or particles with new interactions beyond
the electroweak scale, produced in high-redshift active galactic nuclei.
\end{abstract}

\bigskip



Over the last few years, several giant air showers have been
detected confirming the arrival of cosmic rays (CRs) with energies
greater than a few hundred EeV (1 EeV $\equiv 10^{18}$
eV)\cite{YD}. The nature and origin of these extraordinarily
energetic particles remain a mystery \cite{BS}. The main problem
posed by the detection of CRs of such energy, assuming them to be
photons, nucleons, or nuclei, is that interactions with the
microwave background radiation limit their attenuation length to
less than about 50 Mpc. Therefore, if the CR sources were all at
cosmological distances, the energy spectrum would exhibit the
so-called Geisen-Zatsepin-Kuzmin (GZK)\cite{gzk} cut-off around
80 EeV. Since this is not observed, an astrophysical origin
requires the sources to be within about 100 Mpc. Furthermore,
apart from the energetic difficulties of accelerating particles to
such energies~\cite{acc}, the seeming isotropy on large angular
scales of the observed arrival directions up to the highest
energies~\cite{agasa_clu} leaves only two possibilities for the
source locations: 1) There must be many nearby sources, at least
one close to each arrival direction, but no such convincing source
candidates within 100 Mpc have been found~\cite{ES}. 2) There are
only very few nearby sources which then requires strong
deflection~\cite{mf} in Galactic and/or extragalactic magnetic
fields of micro Gauss strength close to existing upper
limits~\cite{mag}.\\

Recently, Farrar and Biermann \cite{farrar-biermann} have pointed
out the existence of a strong correlation between compact radio
quasars (CRQSOs) and CR events with energies above 80 EeV at
1$\sigma$ level, i.e. events with nominal energies high enough
that the full 1$\sigma$ error bar is above 80 EeV. Specifically,
they have argued that the arrival directions of the CRs of such
energies point back to CRQSOs (redshifts in the range $z = 0.3 -
2.2$) with a probability of chance association of $5 \times
10^{-3}$. If such a correlation is real, it could only be due to
particles generated in these high-redshift sources, which should
traverse unscathed through the primeval radiation evading the GZK
cut-off and being deflected by less than the experimental angular
resolution, of the order of a degree. Note that in such scenarios
the ratio of the signal of neutral particle flux to the charged
particle background depends on many parameters such as the
acceleration process and charged particle deflection by large
scale magnetic fields but should become large above the GZK cut-off.
In the previous analysis the CRQSO-correlation appeared not
to depend strongly on the energy threshold~\cite{hoffman}.\\

Since the energies of the known strongly or electromagnetically
interacting particles drop below $\simeq80\,$EeV during the
propagation from high redshift distances regardless of the initial
energy~\cite{BS}, and since within the Standard Model neutrinos
cannot give rise to the observed showers due to their small
interaction cross section, a clearly established correlation would
most likely indicate new physics. Possibilities involving neutral,
undeflected particles that have been discussed in the literature
include undiscovered neutral hadrons with masses above a few
GeV~\cite{farrar}, and neutrinos attaining cross sections in the
millibarn range above the electroweak scale, which would make them
primary candidates for air showers observed at the highest
energies. Sufficiently heavy neutral particles would avoid pion
production and thus the GZK cut-off, whose threshold energy
increases linearly with rest mass $m$,
$E_{th}=m_\pi(m+m_\pi/2)/\varepsilon$, where $m_\pi$ is the pion
mass and $\varepsilon$ is the background photon energy. Such
particles have been discussed in the context of supersymmetry with
a light gluino, although this possibility appears to be close to
being ruled out~\cite{gluino}. If new physics becomes relevant
around TeV energies, increased neutrino-nucleon cross sections can
occur due to the exchange of graviton Kaluza-Klein modes in the
context of extra dimensions~\cite{extradim} or due to an
exponential increase of the number of degrees of freedom in the
context of string theory~\cite{string}.\\

In the absence of new physics only neutrinos producing
nucleons and photons via resonant Z-production with the relic neutrino
background within about 50 Mpc from the Earth could
give rise to angular correlations with high redshift
sources~\cite{weiler}. However, this requires enormous
neutrino fluxes and/or extreme clustering of relic neutrinos
with masses in the eV range for the interaction rates
to be sufficiently high~\cite{zburst}.\\

Very recently, the Haverah Park experiment presented the analysis
of inclined showers (60$^\circ <$ zenith angle $< 80^\circ$) which
includes two events above 100 EeV \cite{haverah-park}. In
addition, the Akeno Giant Air Shower Array (AGASA) has reported
several remarkable CR events, scattered across half the sky,
\cite{agasa_clu,agasa},  that doubled the original sample used in Ref.
\cite{farrar-biermann}. Thus, and in the light of the theoretical
scenarios mentioned above, it is worthwhile to test again the
possible correlation between the arrival direction of the most
energetic CRs and CRQSOs with flat spectrum. These quasars are
strong radio emitters, a fact that along with their compactness
and variability, is indicative of strong beaming. The bulk of the
observed non-thermal emission of these objects is thought to be
produced in strong, relativistic jets of charged particles emitted
by the active nucleus, which is likely formed by an accreting
supermassive black hole.\\

An interesting sub-group of these sources is formed by the
gamma-ray emitting blazars, which are presumably the most
energetic of them all. There are 66 blazars detected with high
confidence by the EGRET telescope of the Compton Gamma Ray
Observatory, 47 of them in the declination range we are interested
in \cite{3EG}. The 3EG catalog currently contains the most
complete sample of high energy blazars detected so far. Although
the most popular models for gamma-ray emission in these objects
are of leptonic nature, there exists a very interesting family of
hadronic models where the high-energy emission is the result of a
proton-initiated cascade \cite{blaz}. These models open up the
possibility that primaries for ultra-high energy cosmic rays
(UHECRs) above $10^{19}\,$eV could come from secondary reactions
in the hadronic showers, making very energetic EGRET AGN
detections potential candidates for the sources of UHECR events.
We shall use then these 47 EGRET sources as well as the 451 CQSOs
with flat spectrum and declination above $-10^o$ degrees taken
from the surveys of Ref. \cite{kuhr}, to test again the hypothesis
advanced by Farrar and Biermann. We shall use the new and enlarged
AGASA UHECR sample \cite{agasa} plus the highest energy events
detected by Haverah Park \cite{haverah-park,haverah-park2} and
Fly's Eye
\cite{FE}, see Table 1.\\

In order to establish the level of positional coincidence between
QSOs and UHECR events and evaluate its
significance, we shall adopt the code recently developed by Romero
et al. \cite{code} for gamma-ray bursts and unidentified galactic
gamma-ray source studies. This code calculates angular distances
between different kinds of celestial objects in selected catalogs,
and establishes the level of positional correlation between them.
Numerical simulations using large numbers of synthetic populations
(thousands of them were made for each correlation study)  drawn
from an isotropic distribution, i.e. sampled uniformly in right
ascension and declination, are then performed in order to
determine the probability of pure chance spatial association. In
the present case, we generate synthetic populations of the same
number of ultra-high energy cosmic ray events as observed and
compare them with the actual positions of CRQSO and gamma-ray
blazars. We have taken into account firstly that the uncertainties
in the arrival directions of each of the UHECRs is
maintained, i.e. we consider the same positional errors as those
reported for the observed events,  and secondly, that the
artificial sets of UHECR events are constrained (as the actual ones)
to the declination range $\delta
> -10^o$. The treatment of the positional errors is as
follows: we consider a circle around the centroid of each UHECR
event; this circle has a radius equal to the reported 1 sigma
position error for the UHECR. If a CRQSOs or EGRET blazar is
within this circle, we say that there is a positional coincidence.
This procedure was adopted for all events. In the case of the
Fly's Eye and other experiments, where there is an elongated error
box, it was substituted by a circle of similar area. We are not
giving a higher significance to directional coincidences with
small offsets than to coincidences that are not so close, just
because the original errors of the UHECRs are of the order of
degrees. The reader is referred to Ref. \cite{code} for more
details about the procedure.
\\

The results of our analysis are shown in Tables 2 (CRQSOs) and 3
(gamma-ray blazars), where we present, from left to right, the
adopted energy cut-off, the number of real events detected by
AGASA (Ag), Haverah Park (HP), and Fly's Eye (FE), the number of
real positional matches found, the number expected from pure
chance estimated by the simulations, and finally the probability
that the results be the mere effect of chance. In establishing the
positional correlations, both real and simulated, we have adopted
an average error of 1.6$^o$ for the AGASA events, as recommended
in Ref. \cite{agasa}. As a consequence, the highest energy event
of AGASA (Ag213) is not coincident with any CRQSO, contrary to
what was mentioned in Ref. \cite{farrar-biermann}. For the
remaining errors we have kept those used by Farrar and Biermann.
AGASA reports an angular cone radius defined such that in 68\% of
the events, the true direction is contained within the error cone,
it results to be 1.6$^o$ including systematic errors.
Errors in energy for AGASA events were taken as 30\%.\\

\mbox{}From our results using the newest UHECR sample, it can be
seen that the probabilities for the actual coincidence level to be
a random occurrence significantly rise with respect to the
previous work by Farrar and Biermann. The actual coincidences are
all less than 2$\sigma$ away from the simulated mean
value.\footnote{Note also that the UHECR event recorded on
97/03/30, ($E \approx 150$ EeV) which satisfies a restrictive
cut-off energy being at least $\geq 50$ EeV at $2\sigma$ level,
has no CQSO within its error box. Even doubling the error and
searching for background sources with NED, no CQSO appears there.}
In order to test the consistency between our results and those of
Farrar and Biermann for the case of CRQSOs, we repeated the
analysis for the most restrictive cut-off in Table 1 (70 EeV -
$2\sigma$) without taking into account the recent data reported by
Haverah Park, and considering the positional error for Ag213 big
enough for the CRQSO possible counterpart to be included (i.e. an
error of $1.8^o$ as in \cite{1.8}). This situation reproduces the
case reported by Farrar and Biermann (i.e. the event sample
excluding AG110)~\cite{hoffman} and yields a simulated positional
coincidence of 1.75 $\pm$ 0.90, with a chance association
probability of 6~\%, as compared to their number of
1.6\%~\cite{hoffman}. This difference is the result of the use of
a different statistical technique, particularly in the treatment
of positional errors which in our case were taken into account
using top hat functions. We remark, however, that the samples of
both, UHECR events and CRQSOs were the same. Although for the old
data set our analysis method yields chance probabilities larger by
a factor 3$-$4 than theirs, this does not change our main
conclusion, namely that for the new data set the chance
probabilities increase by a factor $> 5$ (within our
analysis) and therefore become insignificant.\\

The correlation with gamma-ray blazars is also likely the result
of chance: we obtain chance probabilities of 26 \% for the highest
energy events and of 46 \% for the events with an energy cut-off
at 27 EeV. For CRQSOs the probabilities are somewhat lower, but
still not significant and significantly above the values given in
Ref.\cite{farrar-biermann}.\\

Virmani et al.~\cite{virmani} recently have also performed a
correlation study. Their analysis shows a remarkable correlation
between UHECRs and CRQSOs, apparently in contradiction with our
result. However, most of their correlation signal comes from
events with large uncertainty both in energy and in position. It
can be seen that independently of the statistical test the
correlation between UHECRs and CRQSOs decreases when considering
only the highest energy events ($E > 8 \times 10^{19}$ eV at
1-standard deviation) that are relevant for new physics because
they have no contamination from the expected proton pile-up around
the photopion production threshold. Furthermore, the QSO sample
used by Virmani et al. is a subsample of ours, formed only by 285
radio loud quasars with flat spectrum obtained from Kuhr's catalog
and checked with NED. Apparently BL LACs or blazars were not
considered, nor undetermined cases. The possibility of the latter
being usual radio galaxies is small because of the flat
spectral index, and consequently both Farrar and Biermann's and
our present study took them into account. Virmani et al. also
included UHECR events from the SUGAR experiment, which is the only
UHECR detector that was operative in the southern hemisphere.
These events strongly contribute to their correlation signal as
can be seen from their Table~1.
However, due to the large detector spacing in SUGAR, their energy
and angular resolution were much poorer than for other experiments
and it is not clear whether the events seen were above the GZK
cut-off~\cite{nwreview}. Finally, the UHECR sample in the
northern hemisphere used by Virmani et al. is different from ours:
we considered 10 UHECR events at most, 8 of them were studied by
Virmani et al., but two recent events from Haverah Park were not.
The positional error in AGASA was 1.6 degrees in our case
\cite{agasa} and 1.8 in theirs. Taking into account these
differences, the statistical methods used by Virmani et al.
would also give a much weaker correlation signal.

In the light of these results, our conclusion is that the
association of CRQSOs and gamma-ray emitting blazars with UHECRs
above the GZK cut-off appears to be not compelling. Hence, there
is currently no support for new multi-GeV neutral hadronic
particles, or for neutrino-nucleon cross sections in the millibarn
range, as explanations of the highest energy cosmic rays, at least
not if these particles are conjectured to be produced in the
classes of sources considered here. We further note that such
scenarios, if there were evidence for them, would require the
sources to accelerate protons at least up to $\sim10^{22}\,$eV,
since the neutral primary candidates have to be produced as
secondaries. While standard acceleration theory requires rather
extreme parameters to achieve that, we note that only a few dozen
such sources in the whole visible Universe would suffice.

\subsection*{Acknowledgments}
This work has been supported by the agencies CONICET and ANPCT
(through grant PICT 98 No. 03-04881), and by Fundaci\'on Antorchas
through separate grants to D.F.T. and G.E.R..

\newpage

\begin{table}[h]
\centering \caption{Cosmic ray events considered in the study.
Errors in position are given, except for the AGASA experiment,
which was considered as a circle of 1.6 degrees radius, see text.
Errors in energy for AGASA events were taken as 30\%, see text.}
\begin{tabular}{clll}
\hline\hline UHECRs  &    energy [$\times 10^{20}$eV] &       RA
(deg) & DEC (deg) \\ \hline FE320  &   3.20 +0.92 -0.94 &    85.2
$\pm$ 0.5 &    48.0 +5.2 -6.3\\ HP120  &   1.20  $\pm$ 0.10 &
179.0 $\pm$ 2.7 &    27 $\pm$ 2.8 \\ HP105 & 1.05 $\pm$ 0.08 &
201.0 $\pm$ 8.7 & 71 $\pm$ 2.5 \\ HP123 & 1.23 +1.0 -0.36 & 86.7
$\pm$ 1 & 31.7 $\pm$ 1.2 \\ HP114 & 1.14 $\pm$ 0.09 &  318.3 $\pm$
1 & 3.0 $\pm$ 2.3
\\ Ag213  &  2.13 & 18.75 & 21.1
\\ Ag144 &    1.44 & 241.5  &               23.0 \\ Ag150 & 1.50
& 294.5 & -05.8 \\ Ag134 &
         1.34  &      280.9 &                48.3 \\
Ag120 &    1.20&                  349               &    12.3 \\
 \hline \hline
\end{tabular}
\label{t0}
\end{table}

\begin{table}[h]
\centering \caption{Positional coincidence (PC), i.e. the number
of real matches within angular resolution, and simulated
positional coincidence (SPC), from an isotropic distribution,
between the highest energy CRs and CRQSOs for different threshold
energies. 27 EeV - $1\sigma$ means, for instance, that the UHECR
events considered have nominal energies such that, subtracting to
it a $1\sigma$ energy error, the result is above 27 EeV.  The last
column indicates the Poisson probability of random occurrence of
any number of coincidences bigger or equal than the real PC.
Columns Ag, HP and FE stand for the number of considered events of
AGASA, Haverah Park and Fly's Eye, respectively.}
\begin{tabular}{ccccccc}
\hline\hline Energy cut-off & Ag & HP & FE & PC & SPC & Prob.\\
\hline $27$ EeV - $1\sigma$ & 58 & - & - &  12 & 8.7 $\pm$ 2.75 &
0.13 \\ $80$ EeV - $1\sigma$ & 5  & 4 & 1 & 4 & 2.7 $\pm$ 1.33 &
0.27 \\ $50$ EeV - $2\sigma$ & 4 & 4 & 1 & 4 & 2.6 $\pm$ 1.28 &
0.26\\ $70$ EeV - $2\sigma$ & 1 & 3 & 1 & 3 & 2.0 $\pm$ 1.01 &
0.31\\ \hline \hline
\end{tabular}
\label{t1}
\end{table}

\begin{table}[h]
\centering \caption{Same as Table I, but for gamma-ray blazars
taken from the Third EGRET catalog.}
\begin{tabular}{ccccccc}
\hline \hline Energy cut-off & Ag & HP & FE & PC & SPC & Prob.\\
\hline $27$ EeV - $1\sigma$ & 58 & - & - & 1  & 0.7 $\pm$ 0.88  &
0.46\\ $80$ EeV - $1\sigma$ & 5 & 4 & 1 & 1 & 0.3 $\pm$ 0.59 &0.26
\\ $50$ EeV - $2\sigma$ & 4 & 4 & 1 & 1 & 0.3 $\pm$ 0.52 & 0.26 \\ $70$ EeV -
$2\sigma$ & 1 & 3 & 1 & 1 & 0.2 $\pm$ 0.47 & 0.19\\ \hline \hline
\end{tabular}
\end{table}

\end{document}